
\font\textfont=cmr12
\font\abstfont=cmr10
\baselineskip=18 pt
\hsize=6.5 truein
\vsize = 9 truein
\textfont
\hfill UAHEP9216
\bigskip
\centerline{\bf TAU DECAY PUZZLES AND POSSIBLE LIGHT GLUINOS$^*$}
\bigskip
\centerline{		   L. CLAVELLI}

\centerline{\it Dept. of Physics and Astronomy}
\centerline{\it    University of Alabama}
\centerline{\it  Tuscaloosa AL 35487 USA}
\bigskip
\centerline{	    Abstract:}
\bigskip
\abstfont\multiply\baselineskip by 2\divide\baselineskip by 3
\parshape= 1 .5in 5.0in
   The relation of the strong coupling constant as measured in tau
   decay to other low energy determinations is considered with
   special attention to its impact on the question of the possible
   existence of light gluinos.
\parskip=12pt
\bigskip
\textfont\multiply\baselineskip by 3\divide\baselineskip by 2
{\bf	      I. The Light Gluino Window}
\medskip
     Although the effects of gluons on experiments in physics have been
present since the measurements of Rutherford in the first decades of
this century, it was only in the late seventies that the existence of
the massless octet of color gauge bosons was firmly established.  Even
to this day the expected existence of new hadronic states with gluon
constituents has not been confirmed although many candidates have been
found.	Examples of scalar mesons which seem to be non-$Q\overline{Q}$
and which may
be gluon-gluon bound states (glueballs) are the $f_0(975)^1$, the
$f_0(1400)^2$, the $f_0(1590)^3$, and the $f_0(1710)^4$.
References
to other observations of these glueball candidates may be found in
ref. 5.

     Given the difficulty of establishing the existence of the glueball
states, it seems likely that other flavor-singlet confined constituents
such as the gluinos of supersymmetry (SUSY) will be first discovered, if
they exist, by indirect effects such as those by which the gluons were
found. Since the squarks must be above half the $Z$ mass due to bounds
from LEP, the primary effects of gluinos are always relatively small
higher order corrections to $QCD$ processes coming from graphs where a
gluon splits into two gluinos. For this reason it is very difficult to
confirm or rule out the presence of light gluinos.  Experimental
constraints are often clouded by model-dependent assumptions related to
hadronization effects, possible \hfill\break
\medskip
\abstfont \multiply\baselineskip by 2\divide\baselineskip by 3
* Talk presented at the Second Workshop on Tau Lepton Physics,
Columbus Ohio, Sept. 8-11, 1992. Proceedings to be published by World
Scientific Press, 1993.\hfil\break
\vfil\eject
\textfont \multiply\baselineskip by 3\divide\baselineskip by 2
\noindent
$R$ parity violation, and to dependence on the photino mass. The hadron
collider limits$^6$
on the gluino mass which
are often quoted as requiring $M_{\tilde G}>150$ GeV
allow, if one reads carefully,
for a gluino mass window below 30 GeV. The presumption is often made
that this window is closed by experiments at $e^+e^-$
colliders.  However
even zero mass gluinos would only perturb the measurements at lepton
colliders through higher order effects and are therefore totally
consistent with current observations.  The CUSB experiment$^7$ suggests
that the gluino mass should not lie between 0.7 and 2 GeV due to the
non-observation of gluinoball-photon final states in $\Upsilon$
decay. This
result, however, is also model dependent in that one has to assume that
the gluinoball wave function at the origin is comparable to that in
quark-antiquark states.  The region below $0.7$ GeV is often thought to
be excluded by the non-observation of $J/\Psi$
decay into this same final
state.
candidates in this region$^{1-4}$
including some seen in $J/\Psi$
decay make it impossible to use charmonium
observations to rule out a light gluino.  If the gluino is below 1
GeV,
the gluinoball states are likely to be strongly mixed with the glueball
states of the same quantum numbers.  The resulting doubling of the
number of non-$Q\overline{Q}$
scalar states is not inconsistent with the number of
available candidates.  Further limits on gluino masses come from beam
dump experiments$^8$
but here again there are significant windows at low
gluino mass.  These are well summarized in figure 5 of ref. 9 which
remains valid at least with regard to gluinos below 0.7 GeV.  Several
low energy windows consistent with those of ref. 9 are also shown in a
study of the relevant data by the UA1 collaboration$^{10}$.
Acknowledgement of the low energy gluino window and attempts to close it
have appeared in the literature from time to time.  In ref. 11, for
example, it was pointed out that the relation between the value of
$\alpha_3$
from $\Upsilon$
decay and that at the $Z$ scale disfavors light gluinos.  In
ref. 12, on the other hand, it was noted that if one considers quarkonia
results taken as a whole there is, in fact, some possible evidence in
favor of light gluinos.  For the most part, however, this light gluino
option which may be the last possible modification of strong interaction
theory in the low energy regime has not received the attention one might
have expected from active researchers in the field.\hfil
\bigskip

{\bf	   II.	The Tau Decay Puzzles}
\medskip
     The gluino can participate in hadronic $\tau$ decay through virtual
effects and, if the gluino mass is below $M_\tau/2$,
through the real decay
 $$\tau \rightarrow \nu_\tau + Q \overline{Q}
				     \tilde{G}\tilde{G}.\eqno(2.1)$$
Both
possibilities are higher order in $\alpha_3$
and therefore small effects.  Such
contributions cannot be addressed until errors are reduced and the
various discrepancies in $\tau$ decay are resolved.  In standard QCD the
electronic branching ratio of the $\tau$,
$B_\tau(e)$, assuming lepton universality
(LU), is given up to cubic order in $\alpha_3$ by
  $$B_\tau(e) = \bigl\lbrace1.973 + 3.0582 \bigl\lbrack1 + \delta_{pert.}
+ \delta_{n.p.}\bigr\rbrack
	\bigr\rbrace^{-1}$$
where$^{13}$
  $$\delta_{pert.} = \left({\alpha_3\over\pi}\right) + 5.2023 \left(
  {\alpha_3\over\pi}\right)^2 + 26.366 \left({\alpha_3\over\pi}\right)
^3 + \cdots\eqno(2.2)$$
\medskip
Non-perturbative QCD effects, $\delta_{n.p.}$, have been estimated$^{14}$
to be less than $1\%$
of the square bracket in eq.(2.2) but these again are somewhat model
dependent.  We prefer to keep model dependent corrections clearly
separate from the fundamental QCD predictions.	Our goal should be to
see how much physics can be predicted from the fundamental theory
relying on model dependent corrections only when needed to resolve clear
(and hopefully small) discrepancies.  Apart from results from lattice
gauge theory which have some claim to rigor, one should not treat
non-perturbative results on the same level as predictions of fundamental
theory.  This will be important in our later discussion.

     Assuming only unitarity and standard model fundamentals such as LU,
$B_\tau(e)$
is strictly proportional to the difference
of the total one-prong branching ratio, $B_1$, and the hadronic one-prong
branching ratio, $B^h_1$.
  $$\lbrack B_1-B_1^h \rbrack/1.973 = B_\tau(e).\eqno(2.3)$$
LU also requires, up to negligible corrections,
   $$\tau_\tau = B_\tau(e) \tau_\mu \left({M_\mu\over M_\tau}\right)^5
		      .\eqno(2.4)$$
Here $\tau_\tau$ and $\tau_\mu$ are the $\tau$ and $\mu$
lifetimes and $M_\tau$ and $M_\mu$ are the
corresponding masses.  An equally strong standard model prediction gives
the $\tau^-$ to $h^-\nu$ decay rate in terms of the
$h^-$  to $\mu^-\nu$ rate.  This relation
for the the sum of the $\tau$ branching ratios to $\pi\nu$ and $K\nu$ is
$$B_\tau(h^-)={\tau_\tau\over{2M^2_\mu M_\tau}}
    \left[{{M_\pi^3 B_\pi(\mu)(M^2_\tau-M^2_\pi)^2}
	  \over{\tau_\pi (M^2_\pi - M^2_\mu)^2}}+{
   {M_K^3 B_K(\mu)(M^2_\tau-M^2_K)^2}\over{\tau_K (M^2_K-M^2_\mu)^2}}
     \right].\eqno(2.5)$$
Thus, given $B_\tau(e)$,
under very mild and well tested assumptions, the left
hand sides of eqs. (2.3) to (2.5) are uniquely predicted.  {\it\/ If}
non-perturbative corrections and as yet unknown higher order QCD
corrections are small, the strong coupling constant, $\alpha_3$,
at the $\tau$ mass
is then determined from eq.(2.2).  It has been pointed out that, because of
the renormalization group convergence, a reliable measurement of the
strong coupling constant in the $\tau$
region is more constraining than a
measurement with the same percentage error at high energy.  Claims have
therefore been made that the $\tau$ data at present provides the best
measurement of $\alpha_3$.
However, the $\tau$ shares the low energy advantage with
several other measurements that have, a priori, equal claims to
reliability including those from deep inelastic scattering, quarkonia
data, lattice calculations, $B\rightarrow J/\Psi + X$,
etc.  If the various measurements
do not yield the same values of $\alpha_3$,
it will take a careful study of
non-perturbative effects to bring them into agreement.	At present the
only rigorous approach to non-perturbative corrections is lattice QCD.
The world average value of $B_\tau(e)$
fit to the known perturbative terms in
eq.(2.2) yields an $\alpha_3(M_\tau)$
value of about 0.3.  This result
contains the tacit assumption that the higher order terms beyond the
cubic are zero.  However, one immediately notes that the quadratic and
cubic terms in eq.(2.2) are each about fifty percent of the preceeding term.
It would seem therefore that a more conservative assumption would be
that the higher order terms continue to decrease in a geometric series.
One would then write

 $$\delta_{pert.} = {\alpha_3\over\pi} [1-5.202{\alpha_3\over\pi}
			 ]^{-1} - 0.69 \left({\alpha_3\over\pi}
				   \right)^3.\eqno(2.6)$$

This expression is totally equivalent to eq.(2.2) up to cubic terms in
$\alpha_3$.
However when fit to the world average value of $B_\tau(e)$ it yields an
$\alpha_3$ about 10\% smaller than the usual procedure.  In table I we give a
range of $\alpha_3$
from zero to 0.36 and the corresponding perturbative
predictions using eq.(2.6) for $B_\tau(e)$
and the three related quantities from
the left hand sides of eqs.(2.3) to (2.5).  On very general grounds the
experimental results should lie on a horizontal line through the last
four columns of table I.  If further higher order and non-perturbative
corrections are small this horizontal line will extend into the first
column yielding the true value of $\alpha_3(M_\tau)$. If the further
corrections are non-negligible the first column will be distorted without
affecting the remaining columns.  The non-perturbative
estimate of reference 14 for example would have the effect of shrinking the
first column from the bottom leading to a larger value of $\alpha_3$
for a given $B_\tau(e)$. The world average values$^5$
of the quantities corresponding to the last four columns of table I are
indicated by the solid vertical lines.	Although the most accurately
known quantity seems to prefer larger values of $\alpha_3(M_\tau)$,
one sees that
any value between zero and 0.33 is consistent with the world average
value of at least one $\tau$
decay property.  Larger values of $\alpha_3$ can also be
obtained if one makes slightly different assumptions concerning the
quartic and higher terms.  The discrepancy between column two and column
four is known as the "one-prong problem".   It is a clear indication
that one or more of the four quantities
$B_\tau(e), \tau_\tau, B_1,$ and $B^h_1$  are being
mismeasured.  The discrepancy between column two and column three is
known as the $\tau$ lifetime problem.  This discrepancy has been somewhat
reduced by the new Beijing measurement of the $\tau$ mass$^{15}$
which has been
confirmed by CLEO and ARGUS.  The errors on the $\tau$ mass are now so small
that a solution of the $\tau$ lifetime problem due to a further change in the
$\tau$ mass is now very unlikely.  The discrepancy between the fifth column
and the second may be called the single hadron discrepancy.  Thus the
world average $\tau$ data shows several independent but interrelated
discrepancies.	Before one can discuss the claim that the $\tau$
provides the best
determination of the strong coupling constant the various discrepancies
should be resolved.  To the extent that lepton universality and unitarity
are not in question table I represents\hfil\break
\smallskip
\abstfont
\baselineskip = 12pt
Table I.\quad The experimental values of the four quantities labeling
columns two through five must lie on a horizontal line assuming only
lepton universality and unitarity. The world average values from ref.
5 are shown in solid lines to the left of each column. The average of
ALEPH and CELLO values are shown in the dotted lines to the right of
of each column.  If non-perturbative effects are small the
corresponding value of $\alpha_3(M_\tau)$ can then be read from the
first column. \hfil\break
\textfont 
\baselineskip = 18pt
\vbox{\tabskip=0pt\offinterlineskip
\def\tablerule{\noalign{\hrule}}
\def\vl{\vrule}
\def\vd{\hbox{:}}
\def\om{\omit}
\halign to 6.5truein{\vrule#\tabskip=6pt plus2pt
minus2pt&\hfil\qquad#\qquad\hfi
  \strut&\vrule#\qquad& #&#&#\qquad& #&#&#\qquad& #&#&#\qquad& #&#&#\qquad&
   \vrule#\tabskip=0pt\cr
\noalign{\vskip4pt}
\tablerule
height2pt&\om&&\om&\om&\om&\om&\om&\om&\om&\om&\om&\om&\om&\om&\cr
 &\hidewidth$\alpha_3(M_\tau)$&&
  \om&\hidewidth$B_\tau(e)$&\om&\om&\hidewidth$\tau_\tau\enskip(ps)$&\om&
   \om&\hidewidth$(B_1-B^h_1)\over{1.973}$&\om&
    \om&\hidewidth$B_\tau(h^-\nu)$&\om&\cr
height4pt&\om&&\om&\om&\om&\om&\om&\om&\om&\om&\om&\om&\om&\om&\cr
\tablerule
height2pt&\om&&\om&\om&\om&\om&\om&\om&\om&\om&\om&\om&\om&\om&\cr
&0.0000&&\om&0.199&\om&\om&324.5&\om&\vl&0.199&\om&\om&0.130&\vd&\cr
&0.0277&&\om&0.198&\om&\om&322.7&\om&\vl&0.198&\om&\om&0.129&\vd&\cr
&0.0554&&\om&0.196&\om&\om&320.7&\om&\vl&0.196&\om&\om&0.128&\vd&\cr
&0.0831&&\om&0.195&\om&\om&318.6&\om&\vl&0.195&\om&\om&0.127&\vd&\cr
&0.1108&&\om&0.194&\om&\om&316.2&\om&\vl&0.194&\om&\vl&0.126&\vd&\cr
&0.1385&&\om&0.192&\om&\om&313.6&\om&\vl&0.192&\om&\vl&0.125&\vd&\cr
&0.1662&&\om&0.190&\om&\vl&310.7&\om&\om&0.190&\om&\vl&0.124&\vd&\cr
&0.1938&&\om&0.188&\vd&\vl&307.5&\om&\om&0.188&\vd&\vl&0.123&\om&\cr
&0.2215&&\om&0.186&\vd&\vl&304.0&\vd&\om&0.186&\vd&\vl&0.121&\om&\cr
&0.2492&&\om&0.184&\vd&\vl&299.9&\vd&\om&0.184&\vd&\vl&0.120&\om&\cr
&0.2769&&\vl&0.181&\vd&\om&295.4&\vd&\om&0.181&\vd&\om&0.118&\om&\cr
&0.3046&&\vl&0.178&\vd&\om&290.1&\vd&\om&0.178&\vd&\om&0.116&\om&\cr
&0.3323&&\vl&0.174&\om&\om&284.0&\vd&\om&0.174&\vd&\om&0.113&\om&\cr
&0.3600&&\om&0.170&\om&\om&276.9&\vd&\om&0.170&\om&\om&0.111&\om&\cr
height2pt&\om&&\om&\om&\om&\om&\om&\om&\om&\om&\om&\om&\om&\om&\cr
\tablerule}}\vfil
\noindent
a purely experimental
problem.  Although their errors are larger than the world average
figures, recent ALEPH and CELLO data suggest a solution in terms of
lower values for columns three and four.  Their combined results are
shown in dotted vertical lines in table I.  A disturbing feature of
their solution however is that they bring down column four at the cost
of increasing various hadronic one prong branching ratios.  Thus their
result for column five tends to increase the single hadron discrepancy.
If future analyses bring this hadronic one-prong rate down again with no
other changes the problem will reappear in column four.

     Although the $\tau$ data has migrated extensively in the past and may
still do so to some extent, new analyses presented at the Dallas
conference in August 1992 and at this meeting tend to suggest an
ultimate horizontal line through table I intercepting $B_\tau(e)$ near 18\%.
The suggested value of $\alpha_3(M_\tau) \approx 0.28$
is much higher than most other low
energy measurements of this coupling constant as can be seen from table
II where each low energy measurement at scale $\mu$ is extrapolated via the
renormalization group to the $\tau$ mass assuming standard particle content.
The average of these values would predict a horizontal line through
table I at $B_\tau(e)\approx 0.187$.  This value is reasonably
consistent with the ALEPH and CELLO results as can be seen from table I. It
is also consistent with a CLEO result published this Summer$^{19}$.  However,
other results, including some announced at the Dallas conference and at this
meeting$^{20}$ suggest preferred values of $\alpha_3(M_\tau)$ as large as
0.36.  Such values may seem pleasing in that they agree after extrapolation
with various measurements at the $Z$. How-
\break
\smallskip
\abstfont
\baselineskip = 12pt
Table II\quad Low Energy determinations of the strong coupling constant
extrapolated to the $\tau$ mass.  The error on the average has been increased
to reflect the scatter in the measurements.  References are given in
parentheses.\hfil\break
\textfont
\baselineskip = 18pt
\vbox{\tabskip=0pt\offinterlineskip
\def\tablerule{\noalign{\hrule}}
\halign to 6.5 truein{\vrule#\tabskip=1em plus2em
   &#\hfil\strut&\hfill#\hfill&\hfill#\hfill&
   \hfill#\hfill&\vrule#\hfil\tabskip=0pt\cr
\noalign{\vskip4pt}
\tablerule
height2pt&\omit&\omit&\omit&\omit&\cr
&\omit&$\mu$\hskip 1em&$\alpha_3(\mu)$&
   $\alpha_3(M_\tau)$&\cr
height4pt&\omit&\omit&\omit&\omit&\cr
&$\Upsilon(1S)\rightarrow GGG$\hskip 1em(12)
	   &4.540&0.181$\pm$0.002&0.250$\pm$0.004&\cr
&$\Upsilon(2S)\rightarrow GGG$\hskip 1em(12)
	   &4.810&0.185$\pm$0.012&0.266$\pm$0.026&\cr
&$\Upsilon(3S)\rightarrow GGG$\hskip 1em(12)
	   &4.970&0.170$\pm$0.007&0.237$\pm$0.014&\cr
&$\Phi(1S)\rightarrow GGG$\hskip 1em(12)
	   &0.420&0.443$\pm$0.007&0.195$\pm$0.001&\cr
&$J/\Psi(1S)\rightarrow GGG$\hskip 1em(12)
	   &1.370&0.191$\pm$0.004&0.176$\pm$0.003&\cr
&$J/\Psi(2S)\rightarrow GGG$\hskip 1em(12)
	   &1.620&0.226$\pm$0.027&0.218$\pm$0.025&\cr
&$\eta_c\rightarrow GG$\hskip 1em(16)
	   &1.490&0.170$\pm$0.030&0.162$\pm$0.027&\cr
&$\Upsilon(1S)\rightarrow \gamma GG$\hskip 1em(16)
	   &1.485&0.230$\pm$0.030&0.215$\pm$0.026&\cr
&$\Upsilon(2S)\rightarrow \gamma GG$ \hskip 1em(16)
	    &1.573&0.200$\pm$0.050&0.192$\pm$0.046&\cr
&$J/\Psi(1S)\rightarrow \gamma GG$ \hskip 1em(16)
	    &0.486&0.430$\pm$0.100&0.201$\pm$0.019&\cr
&Lattice\hskip 1em(17)&5.000&0.174$\pm$0.012&0.246$\pm$0.025&\cr
&Deep Inel.\hskip 1em(18)&10.000&0.156$\pm$0.012&0.272$\pm$0.039&\cr
height4pt&\omit&\omit&\omit&\omit&\cr
&weighted average&\omit&\omit&0.204$\pm$0.027&\cr
height2pt&\omit&\omit&\omit&\omit&\cr
\tablerule}}
\smallskip\noindent
ever one should
keep in mind that the jet measures of $\alpha_3$
are usually strongly dependent
on multi-parameter hadronization monte-carlos.	Such large values of
$\alpha_3$
would have serious negative implications not only for the processes in
table II but also for the prospect of supersymmetric grand unification
(SUSY).  This standard SUSY picture with squarks, sleptons, and gluinos
in the multi-hundred GeV region, which solves the hierarchy problem and
successfully predicts the weak angle and the $b/\tau$
mass ratio, requires an
$\alpha_3(M_Z)$ below 0.114 ($\alpha_3(M_\tau)<0.32$).
Higher values of $\alpha_3$ lead to a SUSY
scale too low to be consistent with CDF bounds on squarks and gluinos
assuming the latter not to be in the low energy window.  The graphical
relation between the SUSY threshold and $\alpha_3(M_Z)$
is shown in figure 4 of
ref. 12 both in the standard picture and in the light gluino case.  In
the light gluino scenario allowable values of $\alpha_3(M_Z)$
run up to 0.118.
Because of the slower running of the strong coupling constant this
corresponds to $\alpha_3(M_\tau)<0.22$.
However, the standard SUSY picture
(non-flipped, heavy gluinos) is already ruled out by the "fine-tuning
problem"$^{21}$ if one requires fine tuning parameters less than 10 and
also by proton decay constraints.  The light gluino may be the only
viable SUSY option (except possibly for flipped SU(5)) and this predicts
a relatively small $\alpha_3(M_\tau)$.

     Apart from this theoretical problem, it is clear from the tables
that if the ultimate horizontal line through the last four columns of
table I intersects $B_\tau(e)$
near or below 0.180, then there are large
non-perturbative effects either in $\tau$ decay or in many of the other low
energy measurements.  The non-perturbative effects in $\tau$ decay have been
estimated to be small $(\delta_{n.p.}$ of eq.(2.2)$\approx$-0.007)$^{14}$.
However one
must keep in mind that, like all non-perturbative calculations except
for lattice QCD, this estimate is highly model dependent.A $B_\tau(e)\approx$
0.18 would be consistent with the average $\alpha_3(M_\tau)$
from table II only if $\delta_{n.p.}$
were larger and positive ($d_{n.p.}\approx$ +0.1).  In fact, $\tau$ decay is
closely related to the $e^+e^-$ hadronic cross section in which there seem
to be large non-perturbative effects at least at charm threshold and in
the low energy region.	If $\alpha_3(M_\tau)$ is as large as 0.3 there is
also a strong scale dependence of the QCD $\Lambda$ parameter deduced from
$\tau$ decay$^{22}$ suggesting again that the perturbation series as now
known does not provide a reliable measure of the strong coupling constant.
For $\alpha_3(M_\tau)\approx 0.2$ this scale dependence essentially disappears.

     If $\alpha_3(M_\tau)$ is large,
one must reconcile table II by
postulating large non-perturbative (relativistic) corrections elsewhere
especially to the
$J/\Psi$ measures$^{23}$.  This is in fact required anyway
if one assumes that the $\Upsilon$
values are more reliable and that there is standard particle content
below the $\Upsilon$.  In such a picture the narrowness of the $J/\Psi$
has little to
do with perturbative QCD since the non-perturbative effects are dominant
and (surprisingly) tend to make the $J/\Psi$ more narrow.  In addition one
then abandons all hope of understanding from perturbative QCD the
narrowness of the $\Phi$ and the Zweig rule as observed at low energy as well
(probably) as the precocious scaling in deep inelastic and the early
appearance of jets at Spear.

     As an alternative approach to the discrepancies of table II, we
have noted that the chi-squared dramatically improves if one postulates
a gluino in the sub-GeV region$^{24}$.	In this scheme the relatively large
apparent values of $\alpha_3$
in the $b$ quark region are due to extra decays of
the bottomonium resonances into gluino containing final states $(GG\tilde{G}
\tilde{G})$.
If one allows for (small) non-perturbative effects by increasing the
errors by 10\% for the $\Phi$ and $J/\Psi$
states and by 5\% for the $\Upsilon$ states, the
best fit with the gluino mass and $\alpha_3(M_Z)$ as free parameters has
 $$M_{\tilde{G}}= 0.44 + .17 GeV \qquad
 \alpha_3(M_Z)= 0.1115^{+.0018}_{-.0013}.\eqno(2.7)$$
Such gluinos fall nicely into the window of refs. 9 and 10.  In addition
to reducing the necessary value of $\alpha_3$
at the $b$ quark scale significantly
and that at the $c$ and $s$ quark scale slightly, the light gluino has the
effect of slowing the running of the strong coupling constant making a
low value at $M_\tau$ consistent with a not-much-lower value at $M_Z$. The
running with the best fit values of eq.(2.7) is shown in fig. 1.  In fig.
1 we show only the values from vector quarkonia together with the
lattice QCD and the deep inelastic result from table II.  These latter
values could conceivably drop slightly with the inclusion of light
gluinos into the analysis.  The effect of light gluinos on the radiative
quarkonia decays has not as yet been analysed but since their errors are
significantly larger it is doubtful that their inclusion will greatly
affect the best fit.  If there are light gluinos the low energy data is
consistent with perturbative QCD and with LEP values of $\alpha_3(M_Z)$
and with
the general SUSY unification picture.  This fit however requires that
$\alpha_3(M_\tau) \approx 0.20$. The slight discrepancy in figure 1 of the
data point from $J/\Psi \rightarrow GGG$ could be removed by adding in an
ad hoc term to represent relativistic corrections as in ref. 23 and the
second of ref. 16.  With light gluinos a much smaller effect would be
attributed to such free parameters than in the treatment of ref. 23.
\vfill
\vbox{
\vskip 3in
\abstfont
Figure 1. Pure QCD fit to quarkonia, deep inelastic, and lattice
data assuming a gluino of mass $0.44$ GeV.\hfil}

     The effect of light gluinos on the $\tau$ has been studied up to second
order in $\alpha_3(M_\tau)^{25}$.
Ignoring the gluino mass, the lowest order effect
is to replace 5.2023 in eq.(2.2) by 4.0647 thus improving the convergence
of the perturbation series.  The cubic contribution however is not as
yet known.  The effect therefore is to increase the value of
$\alpha_3(M_\tau)$
from
$\tau$ decay by about 4.6\% (or less if the gluino effects are suppressed
appreciably by phase space.) In this scenario the low lying hadrons have
some gluino pair components just as the nucleon has a non-vanishing
probability to contain pairs of strange quarks.

     In summary we can distinguish several strong interaction scenarios:
\item{A)}$B_\tau(e)\approx 0.180 ; \quad\delta_{n.p.}$ small in $\tau$ decay
       $(\approx -0.007);\quad \alpha_3(M_\tau)\approx0.3$;\quad
       $\delta_{n.p.}$
large in quarkonia; lattice result problematic; SUSY ruled out
       (except possibly for flipped SU(5)); $\Phi$ decay and Zweig rule
       mysterious

\item{B)}$B_\tau(e)\approx 0.180 ;\quad \delta_{n.p.}$ large in $\tau$
decay $(\approx +0.1); \quad\alpha_3(M_\tau)\approx 0.20$;\quad
 $\delta_{n.p.}$ small in quarkonia; gluinos light; lattice OK; SUSY OK

\item{C)} $B_\tau(e)\approx 0.190;\quad \delta_{n.p.}$ small in $\tau$ decay
 $(\approx-.007?); \quad\alpha_3(M_\tau)\approx0.20$;\quad
 $\delta_{n.p.}$ small in quarkonia; gluinos light, lattice OK; SUSY OK

{}From the point of view of basic theory, alternative C is clearly
preferable since then the fundamental theory predicts the maximum amount
of data.  From this point of view alternative A is least preferable.
Nature of course may not be as tied to perturbation theory as 20th
century theorists, so ultimately experiment must decide.  Apart from the
CLEO 1.5 result$^{19}$ the recent measurements seem to be tending towards
alternative A or B.  Solution A would insure plentiful jobs for
theorists well into the next century in order to resolve the consequent
discrepancies.

     This review was supported in part by DOE grant DE-FG05-84ER40141.
The author acknowledges discussions with collaborators Phil Coulter,
Paul Cox, and Kajia Yuan with whom most of the light gluino
analysis was done.
\bigskip
\centerline{\bf 			    REFERENCES}
\medskip
\baselineskip=12pt
\item{1.}  K.L. Au et al., Phys. Lett. {\it{167B}}, (1984) 229.
\item{2.}  D. Alde et al., Nucl. Phys. {\it{B269}}, (1986) 485.
\item{3.} F. Binon et al., Nuovo Cimento {\it{78A}}, (1983) 313;
Nuovo Cimento {\it{80A}}, (1984) 363.
 \hfil\break  D. Alde et al., Phys. Lett. {\it{B198}}, (1987) 286;
Z. Phys. {\it{C36}}, (1987) 603.\hfil\break
 C. Amsler et al., Phys. Lett. {\it{B291}},(1992) 347.
\item{4.}  J.-E. Augustin et al., Z. Phys. {\it{C36}}, (1987) 369.
   \hfil\break T.A. Armstrong et al., Z. Phys. {\it{C51}}, (1991) 351.
\item{5.}   Particle Data Group, Phys. Rev. {\it{D45}}, No. 11, (1992).
\item{6.}   M. Gold, CDF collab.,talk at Dallas Conference, Aug. 1992.
    \hfil\break  J. Alitti et al., Phys. Lett. {\it{B235}}, (1990) 363.
   \hfil\break F. Abe et al., Phys. Rev. Lett. {\it{62}}, (1989) 1825
\item{7.}   CUSB Collaboration, Phys. Lett. {\it{B156}}, (1987) 233.
\item{8.} F. Bergsma et al., Phys. Lett. {\it{121B}}, (1989) 429.
\item{9.} S. Dawson, E. Eichten, C. Quigg, Phys. Rev. {\it{D31}}, (1985) 1581.
\item{10.}  C. Albajar et al., Phys. Lett. {\it{B262}}, (1987) 109.
\item{11.}I. Antoniadis, J. Ellis, and D.V. Nanopoulos, Phys. Lett. {\it{B262}}
   (1991) 109.
\item{12.}  L. Clavelli, Phys. Rev. {\it{D46}}, (1992) 2112.
\item{13.} S.G. Gorishny, A.L. Kataev and S.A. Larin, Phys. Lett. {\it{B259}},
       (1991) 144.\hfil\break
 L. Surguladze and M. Samuel, Phys. Rev. {\it{D44}}, (1991) 1602.
\item{14.} E. Braaten, S. Narison, and A. Pich, Nucl. Phys. {\it{B373}}, (1992)
      581.
\item{15.} BES Collaboration, talk presented to APS meeting, April 22, 1992.
\item{16.} L. Clavelli, in {\it{High Energy $e^+e^-$ Interactions}},
eds. R.S. Panvini
	  and G.B. Word, AIP Conference Proceedings No. 121, New York,
	  1984.\hfil\break
	  W. Kwong, P. Mackenzie, R. Rosenfeld, and J. Rosner, Phys. Rev.
	  {\it{D37}},(1988) 3210.
\item{17.} A. El-Khadra, G. Hockney, A. Kronfeld, P. Mackenzie, Phys.  Rev.
	  Lett.  {\it{69}}, (1992) 729.
\item{18.}  BCDMS Collaboration, A.C. Benvenuti et al., Phys. Lett. {\it{B237}}
	, (1990) 592; Phys. Lett. {\it{B237}}, (1990) 599.
\item{19.}  CLEO Collaboration, Phys. Rev. {\it{D45}}, (1992) 3976.
\item{20.} K. Gan, CLEO Collaboration, talk presented at the XXVI Int. Conf.
	  on High Energy Physics, Dallas, 1992.\hfil\break
     M. Davier, Summary talk at this Second Workshop on Tau Lepton
	  Physics Columbus Ohio, Sept. 8-11, 1992, published in these
	  proceedings.
\item{21.} R. Barbieri and G. Giudice, Nucl. Phys.  {\it{B306}}, (1988) 63.
\hfil\break P. Nath and R. Arnowitt, Phys. Lett. {\it{B287}}, (1992) 89.
\hfil\break J. Lopez, D. Nanopoulos, A. Zichichi, Phys. Lett. {\it{B291}}
	  , (1992) 255.
\item{22.}  M. Luo and W. Marciano, Hadronic Tau Decays and Perturbative
	Quantum Chromodynamics, preprint INT92-00-01, (1992).
\item{23.}  M. Kobel,
Determination of $\alpha_s$ and $\Lambda$ from Heavy Quarkonium Decays
	  Proceedings of the XXVII Rencontre de Moriond, {\it{QCD and High
Energy
	  Hadronic Interactions}}, ed. Tran Thanh Van, to be published.
\hfil\break  See also the second of ref. 16.
\item{24.} L. Clavelli, P. Coulter, and K. Yuan, Best Fit to the Gluino Mass
	  UAHEP924, to be published.
\item{25.}  L. Clavelli, P. Cox, and K. Yuan, Supersymmetry and $\tau$ Decay,
UAHEP9211, to be published.\vfil
\bye